\documentclass[left=2cm,right=2cm,bottom=2cm,top=2cm]{article}
\usepackage{amsmath,amssymb}
\usepackage{graphicx}
\usepackage{dcolumn}
\usepackage{bm}
\usepackage{hyperref}

\begin{document}

\title{Vanishing Dimensions in Four Dimensional Cosmology with Nonminimal Derivative Coupling of Scalar Field}

\author{Agus Suroso and Freddy P. Zen\\
Theoretical Physics Laboratory, THEPI Division, and \\
INDONESIA Center for Theoretical and Mathematical Physics (ICTMP)\\
Faculty of Mathematics and Natural Sciences, Institut Teknologi Bandung \\
Jl. Ganesha 10 Bandung 40132, Indonesia.\\
email: agussuroso@fi.itb.ac.id, fpzen@fi.itb.ac.id\\
}%



\maketitle

\begin{abstract}
We discussed a mechanism that allows the universe to start from lower dimension ($d < 4$) in its very early era and evolves to four dimension at the end of the process. The mechanism is generated by a  nonminimal derivative coupling of scalar field, where derivative terms of the scalar field coupled to curvature tensor. We solve the Einstein equations for a limit of large and nearly constant Hubble parameter and analyze the evolution 
of the Hubble parameter in this limit.
 The lower-dimensionality in early universe has advantages in the context of unification theory. 
\end{abstract}


Physics of early universe has attracted much attention. One of the main issue is the number of dimensions of the early universe, which is strongly related to the unification of all physical theories.
String theory, as a candidate for the unification theory of physics, suggests that the universe consists of 26 (for bosonic string) and 10 dimensions (for superstring). As the universe evolves,  the extradimensions compactified to a very small size, too small to be observabed.
The idea of compactified extra dimensions then developed into braneworld model of gravity. In braneworld model, our four dimensional spacetime is embedded in five dimensional bulk and the matter field is constrained in the brane while the graviton can move freely in the bulk.
One of the most popular model of braneworld scenario is the Randall-Sundrum model \cite{RSI,RSII}. 

Another candidate of unified theory of all physical phenomena is the quantum gravity. This theory has a good mathematical consistency in 1+1 and 1+2 dimensions, but not in 1+3 dimension (\cite{Carlip:1998uc}, and reference therein), which is the number of  observable dimensions. Some attempts has been done to extend the theory to 1+3 dimensions. There is a model proposed in \cite{jonas} can be used to solve this problem. They suggest that dimensionality of the universe depends on the scale of observation, so that in a small scale the universe has smaller dimension (less than 1+3) and in a large scale the universe may have a large dimension (may have extradimensions). 

In this letter, we propose another approach concerning the dimensionality of the universe. Instead of the extradimension of early universe compatified or scale dependent dimensions, we suggest that universe has evolved from small dimension
in very early universe to 1+3 dimensions now. We adopt the idea of inhomogeneous universe proposed in \cite{tomita, Watanabe:2009ct}. 
In inhomogeneous universe, one spatial dimension can evolve faster or slower than other spatial dimensions.

A spatially inhomogeneous early universe is described by the following \emph{ansatz} metric, 
\begin{equation}
  ds^2 = -dt^2 + a_1^2 dx^2 + a_2^2 dy^2 + a_3^2 dz^2,
\end{equation} 
where $a_i=a_i(t)$, $i=1,2,3$  is scale factors for every spatial coordinate. When $a_1=a_2=a_3$ the universe will be spatially homogeneous.
Furthermore, we can describe a  $1+1$ dimensional universe by choosing two of the scale factors zero or small enough comparied to another one,  and get $1+2$ dimensional universe where one of the scale factors is zero. The geometrical quantities that derived from the metric are
\begin{align}
  R_{00} &= - \left(\dot{H}_1 + \dot{H}_2 + \dot{H}_3 + H_1^2 +H_2^2 +H_3^2 \right),\\
  R_{11} &= a_1^2 \left(\dot{H}_1 + H_1^2 + H_1 H_2 + H_1 H_3 \right),\\
  R_{22} &= a_2^2\left(\dot{H}_2 + H_2^2 + H_1 H_2 + H_2 H_3 \right),\\
  R_{33} &= a_3^2 \left(\dot{H}_3 + H_1^3 + H_1 H_3 + H_2 H_3 \right),\\
  R &= 2 \left(\dot{H}_1 + \dot{H}_2 + \dot{H}_3 + H_1^2 +H_2^2+H_3^2 + H_1 H_2 + H_2 H_3 + H_1 H_3 \right),\\ 
  G_{00} &= H_1 H_2 + H_1 H_3 +H_2H_3,\\
  G_{11} &= -a_1^2 \left(\dot{H}_2 + \dot{H}_3+ H_2^2 + H_3^2 +H_2 H_3 \right),\\
  G_{22} &= -a_2^2 \left(\dot{H}_1 + \dot{H}_3+ H_1^2 + H_3^2 +H_1 H_3 \right),\\
  G_{11} &= -a_1^2 \left(\dot{H}_1 + \dot{H}_2+ H_1^2 + H_2^2 +H_1 H_2 \right).
\end{align}

As the dynamical quantity that makes the spacetime evolves, we introduce the nonminimal derivative coupling of scalar field in following action,
\begin{align}
\label{eq:aksi}
S=\int d^4x \sqrt{-g} \left(\frac{R}{2\kappa^2}  + \mathcal{L}(\phi)\right),
\end{align}
where
\begin{equation}
\mathcal{L}(\phi)= \frac{1}{2}g_{\mu\nu} \partial^\mu \phi \partial^\nu \phi + \frac{\xi}{2} R g_{\mu\nu} \partial^\mu \phi \partial^\nu \phi+ \frac{\eta}{2} R_{\mu\nu}  {\partial^\mu \phi \partial^\nu \phi}.
\end{equation}
This coupling was introduced in \cite{amendola} and has been studied extensively by some authors in the context of four \cite{amendola, Capozziello:1999xt, capozziello2,granda,sushkov} and five dimensional gravity \cite{Suroso2013799, Suroso:2012tda,Suroso:2011zz}. 
In \cite{Suroso2013799} we assume that in its very early era, our universe consisted of five dimensional spacetime then the four dimensional and one dimensional extra dimension evolved differently. We find a solution for recent universe consisting of expanding four dimensional spacetime  with acceleration and shringking extra dimension  is exists. The solution for that case is the de Sitter universe where the Hubble constant evolves from a large value to a small one.

Now, let us consider an inhomogeneous universe with $a_1 = a_2 = a$ and $a_3 = a^\gamma$, where $\gamma$ is a non-constant parameter.
For every value of $a$, the condition of  $\gamma < 0$ will give us a 1+3 dimensional universe with one dimension (which is related to the scale factor $a_3$) smaller than two other spatial dimensions. In other words, we have a 1+2 dimensional universe.
On the other hand, condition of  $\gamma >0$ will make  $a_3$ dominant on other two factors and the universe become 1+1 dimensional. 
Then the condition of $\gamma = 1$ will give us a homogenous 1+3 dimensional universe. In general, we can choose
 $\gamma$ to depend on time,
$\gamma = \gamma(t)$. With this choice, we can define Hubble parameters for each coordinates as follows,
\begin{align}
  H_1 &= H_2 \equiv H,\\
  H_3 &= \dot{\gamma} \ln a + \gamma H\\
  \dot{H}_3 &= \ddot{\gamma}\ln a + \dot{\gamma} H + \gamma \dot{H}.
\end{align}
 
In general, the field equations that comes from the action (\ref{eq:aksi}) contain some higher order terms of time derivative of the scalar field. As given in \cite{Suroso2013799, sushkov}, the condition of $2\xi +\eta =0$ will remove the higher order terms. In this paper, we do choose that condition. The Einstein equations that comes from action (\ref{eq:aksi}) are
\begin{align}
  \left( 1- \xi\kappa^2\dot{\phi}^2\right)\left(2\gamma +1\right)H^2 &= \frac{\kappa^2\dot{\phi}^2}{2},\\
  \label{eq:einstein11}
  G_{11}\left(1+\xi\kappa\dot{\phi}^2\right) &= g_{11}\left(\frac{1}{2} + 2R_{00}\right)\dot{\phi}^2,\\
  \label{eq:einstein33}
  G_{33}\left(1+\xi\kappa\dot{\phi}^2\right) &= g_{33}\left(\frac{1}{2} + 2R_{00}\right)\dot{\phi}^2.
\end{align}
We do not write the equation for 22-component because its value is the same as that of  11-component. Dividing (\ref{eq:einstein11}) with  (\ref{eq:einstein33}) gives
\begin{align}
  \left(\gamma-1\right) \dot{H} &+ \left(\gamma^2+\gamma-2\right) H^2 + \left(\dot{\gamma} + 2\dot{\gamma}\gamma \ln a + \dot{\gamma}\ln a\right) H\nonumber\\ 
\label{eq:11bagi33}
  &+ \ddot{\gamma} \ln a + \dot{\gamma}^2 (\ln a)^2 = 0.
\end{align} 
Next, we consider to solve this equation to find  the value of the $\gamma$ parameter.

As the simplest solution, we consider the de Sitter solution with large and nearly constant Hubble parameter. For this solution, we take $H=H_0$ as a constant and eq. (\ref{eq:11bagi33}) can be rewritten as 
\begin{equation}
\left[ \dot{\gamma}^2t^2 + \left(2\gamma+1\right) \dot{\gamma}t + \gamma^2 + \gamma -2 \right] H_0^2   =0
\end{equation}
Then we have the following conditions,
\begin{align}
\dot{\gamma}^2t^2 + \left(2\gamma+1\right) \dot{\gamma}t + \gamma^2 + \gamma -2 
=\left[ \dot{\gamma}t+\left( \gamma +2\right)\right]\left[ \dot{\gamma}t+\left( \gamma -1\right)\right]= 0,\\
\end{align}
and the solutions are
\begin{align}
  \gamma = -2 + \tilde{\gamma}_1 e^{-t},
\end{align}
or
\begin{align}
\label{eq:gamma2}
  \gamma = 1 + \tilde{\gamma}_2 e^{-t},
\end{align}
where $\tilde{\gamma}_1$ and $\tilde{\gamma}_2$  areintegration constants. If we choose $\tilde{\gamma}_1 = 3$, the value of $\gamma$
varies from $+1$ when $t\rightarrow 0$ to $-2$ when $t\rightarrow \infty$. It means that the universe evolve from $1+3$ dimensions to $1+2$ dimensions. Then if we choose $\tilde{\gamma}_2 =-2$, the value of $\gamma$ varies from $-1$ when $t \rightarrow 0$ to $+1$ when $t\rightarrow\infty$, which means that the universe evolve from $1+2$ dimensions to $1+3$ dimensions. It seems that eq. (\ref{eq:gamma2}) is more natural solution than the other one. 

According to solution (\ref{eq:gamma2}), after long enough periods of time the gamma value become constant and close to +1. When $\gamma$ is constant, eq. (\ref{eq:11bagi33}) give us
\begin{equation}
  \dot{H} + \left(\gamma +2\right)H^2=0,
\end{equation}
and the solution is
\begin{equation}
\label{eq:aakhir}
   a \propto \left[ \left(\gamma +2\right)\left(t-t_0\right)\right]^{\frac{1}{\gamma +2}}.
\end{equation}
Then for $\gamma = 1$ we have $a = c \propto (t-t_0)^{1/3}$ which is equal to the case of universe that is dominated by stiff matter. Hence our model show that soon after inflation (de Sitter era)  universe was dominated by stiff matter. The model of early universe with stiff matter has been studied in \cite{neto} and references therein. 
 
\begin{figure}
  \includegraphics[width=10cm]{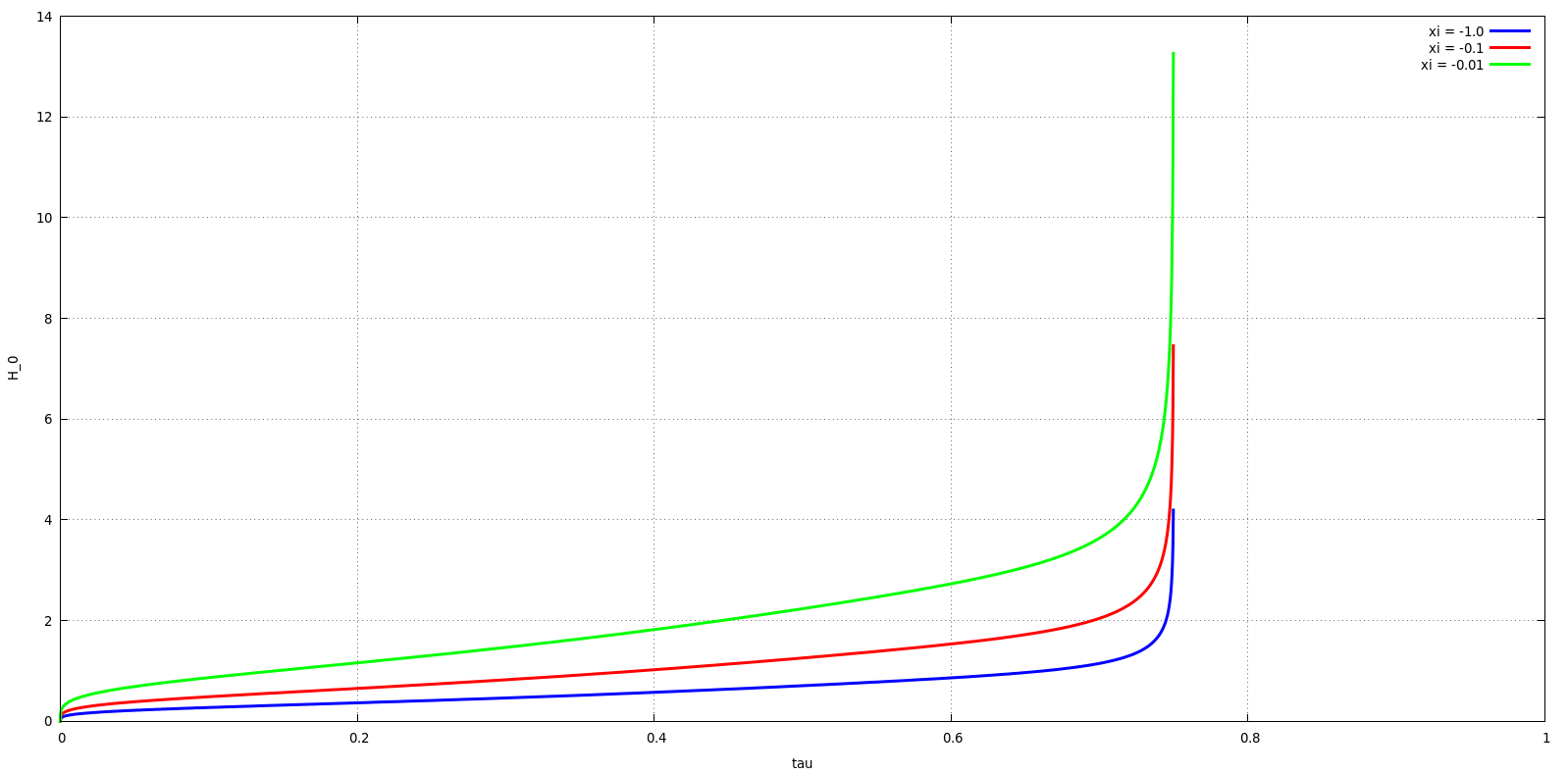}
  \includegraphics[width=10cm]{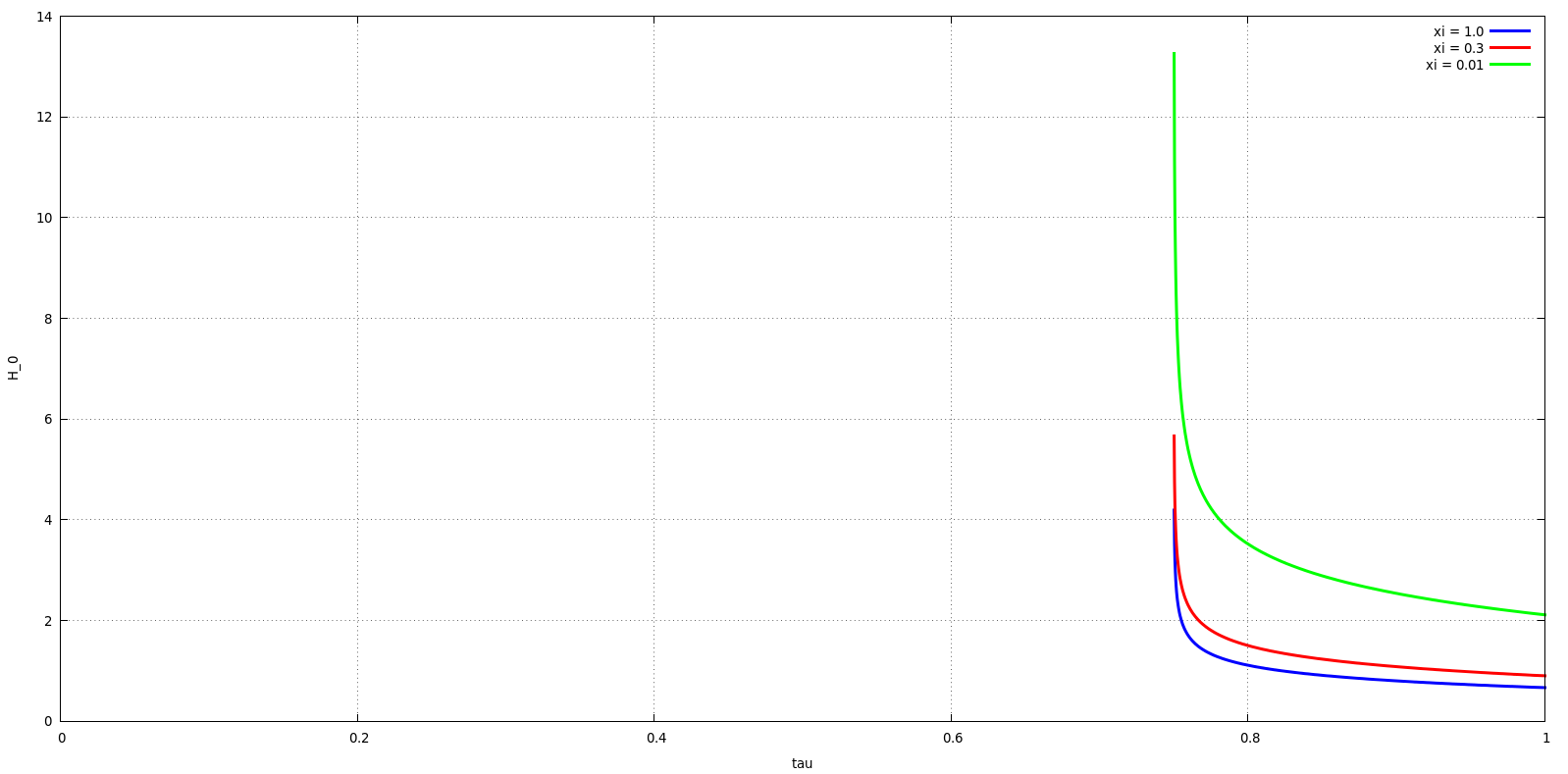}  
  \caption{Plots for $H_0$ vs $\tau$ for de Sitter solution with  nearly constant Hubble parameter.}.
  \label{fig:HvsTau}
\end{figure}

As consistency condition, we check the equation of motion of the scalar field,
\begin{equation}
  -\ddot{\phi} - \left(2 + \gamma\right)H \dot{\phi} - 2\xi \left(G^{00}\ddot{\phi} -2a_1^2 H^2 G^{11}\dot{\phi} - \gamma^2 H^2 a_3^2G^{33}\dot{\phi} \right) = 0.
\end{equation} 
For de Sitter solution of large and nearly constant Hubble parameter, the equation of motion can be reduced to
\begin{equation}
  - \frac{2\dot{\gamma}}{4\gamma+2} = 2\xi \left( 5\gamma^2+2\gamma+2\right) H_0^4.
\end{equation}
Substituting eq. (\ref{eq:gamma2}) for $\gamma$, choosing $\tilde{\gamma}_2 = -2$, and writing $t = -\ln(\tau)$, the last equation can be rewritten as
\begin{equation}
 −\frac{ \tau}{2\left( 1-2\tau\right) +1}=\left[ 2\left( 1-2\tau\right) +5{\left( 1-2\tau\right) }^{2}+2\right] \xi{H_0}^{4}
\end{equation}
The plots for $H_0$ vs $\tau$ for negative and positive value of $\xi$ are given in figure \ref{fig:HvsTau}.
From the figure we see that for $\xi <0$ the Hubble parameter dissapears (or has not appeared yet) at early time ($\tau>0.75$), and begin to evolve from a  maximum value at $\tau \rightarrow 0.75$ to zero value at $\tau \rightarrow 0 $ (or $t \rightarrow \infty$). On the other hand, for $\xi>0$ the Hubble parameter increases from a particular value at $t=0$ to a maximum value at $\tau \rightarrow 0.75$ and then dissapear.  For both cases, increasing magnitude of $\xi$ will increase the maximum value of Hubble parameter. 

The case of $\xi<0$ have more interesting features than $\xi>0$. We can make a speculation that the evolution of Hubble parameter for the case is related to inflation of the universe, where the limit of $\tau \rightarrow 0.75$ is related to the era when the inflation begin. At the end of the process ($\tau \rightarrow 0$), the Hubble parameter goes to zero, which is consistent with the result of eq. (\ref{eq:aakhir}), and the $\gamma$ parameter goes to 1 (or the spacetime goes to four dimensional). 

As conclusion, we discussed a mechanism that allows the universe to start from lower dimensional ($d < 4$) in its very early era and evolve to four dimension at the end of the process. The mechanism is generated by a  nonminimal derivative coupling of scalar field, where derivative terms of the scalar field is coupled to curvature tensor. The lower-dimensionality in early universe has advantages in the context of unification theory. 
It is believed that at very early era when the universe has very high energy, the quantum and gravity phenomena described by quantum gravity  works only in 1+2 dimensions.

\textbf{Acknowledgement}. {The work of AS was supported by Program Riset dan Inovasi Kelompok Keilmuan ITB 2014. The work of FPZ was supported by Program Riset dan Inovasi Kelompok Keilmuan ITB 2014 and Program Riset Desentralisasi DIKTI 2014. We thank Jusak S. Kosasih for careful reading and correcting English grammar.




\end{document}